\documentclass [11pt]{article}
\title{Vacuum Fluctuations Cannot Mimic a Cosmological Constant}
\author{Robert D. Klauber\\1100 University Manor Dr., 38B \\Fairfield, Iowa 52556
\\rklauber(AT)iowatelecom.net, permanent: rklauber(AT)netscape.net}

\date{Original 19 July 2007; minor revision 31 Oct 2007}

\usepackage{graphicx}
\usepackage{longtable}

\begin{document}

\maketitle

\begin{abstract}

When the vacuum fluctuation pressure is calculated directly from fundamental 
principles of quantum field theory, in the same manner as vacuum fluctuation 
energy density is commonly calculated, one finds it is not equal to the 
negative of the vacuum fluctuation energy density. Thus, vacuum fluctuations 
cannot manifest as a cosmological constant of any order.

\end{abstract}

\section{Introduction}
As summarized by Peebles and Ratra \cite{Peebles:2003}, Padmanabhan 
\cite{Padmanabhan:2003}, and others, there are presently three overriding 
cosmologic issues involving phenomena for which no generally accepted 
theoretical solutions exist: 1) dark matter (non-baryonic, unseen ``normal'' 
matter), 2) dark energy (small positive cosmological constant or 
quintessence), and 3) a vanishing sum of zero-point energies.

We will not deal with 1) here. In \cite{Klauber:1}, I show a possible 
mechanism for 3), somewhat related to ideas proposed by others 
\cite{Linde:1988} - \cite{Hooft:2006}.
In the course of investigating 2), many of these authors, as well as many others, 
such as \cite{Weinberg:1989} - \cite{Ishak:2007}, have 
surmised that the zero-point energy which arises from the so-called vacuum 
fluctuations leads to a cosmological constant, albeit one that, by the most 
straightforward quantum field theory (QFT) calculations, differs from the 
observed value by a factor of almost 10$^{120}$ \cite{Weinberg:1989}.

In this article, I show that the vacuum fluctuations, while they give rise 
to vacuum energy, do not give rise to the appropriate vacuum pressure needed 
to result in a cosmological constant of any magnitude.

\section{Background}
\label{sec:background}
\subsection{Lorentz Invariance of the Vacuum}
\label{subsec:lorentz}
The stress energy tensor for a perfect fluid, which the vacuum is assumed to 
be, has the form in the local rest mass frame of
\begin{equation}
\label{eq1}
\left\langle {T^{\mu \nu }} \right\rangle \,=\,\left[ 
{{\begin{array}{*{20}c}
 \rho \hfill & 0 \hfill & 0 \hfill & 0 \hfill \\
 0 \hfill & p \hfill & 0 \hfill & 0 \hfill \\
 0 \hfill & 0 \hfill & p \hfill & 0 \hfill \\
 0 \hfill & 0 \hfill & 0 \hfill & p \hfill \\
\end{array} }} \right].
\end{equation}
Most consider that the stress-energy tensor for the vacuum should be Lorentz 
invariant, which means that it must take the form
\begin{equation}
\label{eq2}
\left\langle {T_{\mbox{vac}}^{\mu \nu } } \right\rangle \,=\,\rho 
_{\mbox{vac}} \,\eta ^{\mu \nu }=\rho _{\mbox{vac}} \left[ 
{{\begin{array}{*{20}c}
 1 \hfill & 0 \hfill & 0 \hfill & 0 \hfill \\
 0 \hfill & {-1} \hfill & 0 \hfill & 0 \hfill \\
 0 \hfill & 0 \hfill & {-1} \hfill & 0 \hfill \\
 0 \hfill & 0 \hfill & 0 \hfill & {-1} \hfill \\
\end{array} }} \right]\quad ,
\end{equation}
and hence,
\begin{equation}
\label{eq3}
p_{\mbox{vac}} =-\rho _{\mbox{vac}} \quad .
\end{equation}
Expressing the equation of state for any perfect fluid as
\begin{equation}
\label{eq4}
p=w\rho \quad ,
\end{equation}
for the vacuum, we would have
\begin{equation}
\label{eq5}
w_{\mbox{vac}} =-1\quad .
\end{equation}
\subsection{Effective Cosmological Constant}
\label{subsec:effective}
In the Einstein field equations, the cosmological constant of scalar value 
$\Lambda $ must take the tensor form
\begin{equation}
\label{eq6}
\Lambda \eta ^{\mu \nu }\quad .
\end{equation}
Thus, from (\ref{eq1}), any perfect fluid which might mimic a cosmological constant 
must have
\begin{equation}
\label{eq7}
w_{\mbox{perf}\,\mbox{fluid}\,\mbox{CC}} =-1
\end{equation}
\subsection{Constant Vacuum Energy Density}
\label{subsec:constant}
It is well known \cite{Peebles:1993}, and derivable from the Einstein field 
equations with a Friedmann metric, that the rate of change of energy density 
for a perfect fluid in an expanding universe is
\begin{equation}
\label{eq8}
\dot {\rho }=-3\left( {\rho +p} \right)\frac{\dot {a}}{a}
\end{equation}
where $a$ is the time dependent scale factor of the universe (the ``radius'' of 
the universe). One would expect the energy density of the vacuum to remain 
constant, and from (\ref{eq8}), it can only do so if the vacuum pressure is equal in 
magnitude and opposite in sign from the vacuum density, i.e., as in (\ref{eq5}).

\subsection{Cosmological Observation}
\label{subsec:cosmological}
Recent observation \cite{Reis:1}\cite{Schwarzschild:2007} indicates that
\begin{equation}
\label{eq9}
w_{\mbox{vac}} =-1\,\,\,\,\left( {\,\pm 10\% } \right)\quad ,
\end{equation}
and the data suggests that this value has been constant, or very close to 
constant, for at least, a substantial part of the history of the universe. 
Thus, these observations support the existence of a cosmological constant. 
Further, the observations indicate it has a small, positive value, leading 
to accelerating expansion of the universe.

\subsection{Kinds of Vacuum Energy}
There are primarily two known ways in which the vacuum can possess energy, 
a) vacuum fluctuations and b) symmetry breaking remnants. Both have been 
proposed as possible sources for the observed cosmological constant.

\subsubsection{Vacuum fluctuations}
In second quantization, one takes the classical field theory Poisson 
brackets over into commutators. Thus, the field (e.g. \textit{$\phi $}) and its conjugate 
momentum (e.g. $\pi )$ do not commute, and this leads to a Hamiltonian 
operator with an infinite series of $\raise.5ex\hbox{$\scriptstyle 
1$}\kern-.1em/ \kern-.15em\lower.25ex\hbox{$\scriptstyle 2$} $ quanta terms. 
The vacuum expectation value for energy then becomes infinite (or finite, 
but extremely large, if a suitable energy cutoff, such as the Planck energy, 
is employed). \cite{Weinberg:1989}

\subsubsection{Symmetry breaking remnants}
In the typical Higgs symmetry breaking mechanism, particles gain mass as the 
universe cools, and the vacuum gains an energy density (the remnant). This 
is simply a constant term arising in the Hamiltonian density after the 
symmetry breaking, usually designated as $V$. This remnant energy density is 
typically on the order of the energy density of the mass-energy density in 
the universe at the time of the symmetry breaking. This is far smaller than 
that calculated for the vacuum fluctuations, yet far larger than that 
observed.

\subsection{Conclusions for the Vacuum}
Lorentz invariance, constancy of vacuum energy density, and observation all 
indicate a vacuum pressure which is negative and equal in magnitude to the 
vacuum energy density. Both vacuum fluctuations and symmetry breaking 
remnant energy densities are far larger than what is observed, but various 
corrective mechanisms have been proposed for both, which presumably might 
make either a candidate for the observed cosmological constant.

\section{Determining Vacuum Pressure from Theory}
\subsection{The Common Approach}
Many researchers \cite{Volovik:1} use vacuum energy density relations, 
expressed in the form of (\ref{eq8}) or its sibling thermodynamic relation 
(derivable from (\ref{eq8})),
\begin{equation}
\label{eq10}
d(\rho V)=-pdV\quad ,
\end{equation}
where $V$ here is volume, to deduce (\ref{eq3}). Further justification is derived from 
the Lorentz invariance logic of Section \ref{subsec:lorentz}.

This is then used to justify that both
\begin{equation}
\label{eq11}
p_{\mbox{vac}\,\,\mbox{flucts}} =-\rho _{\mbox{vac}\,\,\mbox{flucts}} 
\end{equation}
and
\begin{equation}
\label{eq12}
p_{\mbox{sym}\,\,\mbox{remnant}} =-\rho _{\mbox{sym}\,\,\mbox{remnant}} 
\quad .
\end{equation}
\subsection{The Fundamental Theoretic Approach}
\label{subsec:mylabel1}
For the symmetry breaking remnant, Peebles \cite{Ref:1} takes a more 
fundamental approach, which does not assume that the vacuum energy densities 
we calculate from theory must, \textit{a priori}, have $w$ = -- 1.

Peebles starts with the stress-energy tensor for a quantum field and finds 
the energy density and the pressure in the vacuum from that field. From 
this, he proves that the symmetry breaking remnant, under the proper 
conditions, can give rise to a $w_{\mbox{sym}\,\,\mbox{remnant}} =-1$.

Critically, Peebles does not simply assume that pressure from a vacuum 
contribution must equal the negative of the energy density. From fundamental 
principles of QFT, he proves it.

\subsection{Lack of Application of Fundamental Approach to Vacuum Fluctuations}
\label{subsec:mylabel2}
No one known to this author applies Peebles fundamental theoretic approach 
to vacuum fluctuations, the other form of vacuum energy density. Further, 
the assumption that (\ref{eq11}) holds is widespread, in fact dominant, in the 
literature.

In the next section, we apply the fundamental approach of Peebles to vacuum 
fluctuations, to determine the true relation between pressure and energy 
density of the vacuum for these fluctuations \cite{See:1}.

\section{Determining Pressure of Vacuum Fluctuations}
For simplicity, we restrict ourselves to the free real scalar field, with 
Lagrangian density
\begin{equation}
\label{eq13}
{\cal L}_\phi =\textstyle{1 \over 2}\left( {\partial _\mu \phi g^{\mu \nu 
}\partial _\nu \phi -m^2\phi ^2} \right)\quad .
\end{equation}
It can be shown \cite{Ref:1} that
\begin{equation}
\label{eq14}
T_\phi ^{00} =\rho _\phi =\textstyle{1 \over 2}\left( {\left( {\dot {\phi }} 
\right)^2+\left( {\nabla \phi } \right)^2+m^2\phi ^2} \right)
\end{equation}
\begin{equation}
\label{eq15}
T_\phi ^{11} =p_\phi =\left( {\partial _1 \phi } \right)^2+\textstyle{1 
\over 2}\left( {\left( {\dot {\phi }} \right)^2-\left( {\nabla \phi } 
\right)^2-m^2\phi ^2} \right).
\end{equation}
First, we review the well known calculation of (\ref{eq14}), and then follow similar 
steps to determine (\ref{eq15}).

\subsection{Energy Density of Vacuum Fluctuations}
\label{subsec:energy}
The real scalar field and its derivatives are
\begin{equation}
\label{eq16}
\phi =\sum\limits_{\rm {\bf k}} {\frac{1}{\sqrt {2V_s \omega _{\rm {\bf k}} 
} }\left( {a({\rm {\bf k}})e^{-ikx}+a^\dag ({\rm {\bf k}})e^{ikx}} \right)} 
\end{equation}
\begin{equation}
\label{eq17}
\dot {\phi }=\sum\limits_{\rm {\bf k}} {\frac{i\omega _{\rm {\bf k}} }{\sqrt 
{2V_s \omega _{\rm {\bf k}} } }\left( {-a({\rm {\bf k}})e^{-ikx}+a^\dag 
({\rm {\bf k}})e^{ikx}} \right)} 
\end{equation}
\begin{equation}
\label{eq18}
\phi ,_i =\sum\limits_{\rm {\bf k}} {\frac{ik_i }{\sqrt {2V_s \omega _{\rm 
{\bf k}} } }\left( {a({\rm {\bf k}})e^{-ikx}-a^\dag ({\rm {\bf k}})e^{ikx}} 
\right)} ,
\end{equation}
where $V_{s}$ is spatial volume. The first term on the RH of (\ref{eq14}) is
\begin{equation}
\label{eq19}
\textstyle{1 \over 2}\dot {\phi }\dot {\phi }=\textstyle{1 \over 2}\left( 
{\sum\limits_{\rm {\bf k}} {\frac{i\omega _{\rm {\bf k}} }{\sqrt {2V_s 
\omega _{\rm {\bf k}} } }\left( {-a({\rm {\bf k}})e^{-ikx}+a^\dag ({\rm {\bf 
k}})e^{ikx}} \right)} } \right)\left( {\sum\limits_{{\rm {\bf {k}'}}} 
{\frac{i\omega _{{\rm {\bf {k}'}}} }{\sqrt {2V_s \omega _{{\rm {\bf {k}'}}} 
} }\left( {-a({\rm {\bf {k}'}})e^{-i{k}'x}+a^\dag ({\rm {\bf 
{k}'}})e^{i{k}'x}} \right)} } \right).
\end{equation}
In
\begin{equation}
\label{eq20}
\langle \phi _{\rm {\bf k}} \vert \rho _\phi \vert \phi _{\rm {\bf k}} 
\rangle 
\end{equation}
where $\rho _\phi $ is an operator represented by (\ref{eq14}), and $\vert $\textit{$\phi $}$_{k}>$ 
can be any state, all terms with ${\rm {\bf k}}\ne {\rm {\bf {k}'}}$ drop 
out, as do all terms in $a({\rm {\bf k}})a({\rm {\bf k}})$ and $a^\dag ({\rm 
{\bf k}})a^\dag ({\rm {\bf k}})$. (\ref{eq19}) is part of (\ref{eq14}), so using the 
commutation relations
\begin{equation}
\label{eq21}
\left[ {a({\rm {\bf k}}),a^\dag ({\rm {\bf {k}'}})} \right]=\delta _{{\rm 
{\bf k{k}'}}} 
\end{equation}
that part reduces to
\begin{equation}
\label{eq22}
\textstyle{1 \over 2}\dot {\phi }\dot {\phi }\to \textstyle{1 \over 
2}\sum\limits_{\rm {\bf k}} {\frac{\left( {\omega _{\rm {\bf k}} } 
\right)^2}{2V_s \omega _{\rm {\bf k}} }\left( {a({\rm {\bf k}})a^\dag ({\rm 
{\bf k}})+a^\dag ({\rm {\bf k}})a({\rm {\bf k}})} \right)} =\frac{1}{2V_s 
}\sum\limits_{\rm {\bf k}} {\omega _{\rm {\bf k}} \left( {a^\dag ({\rm {\bf 
k}})a({\rm {\bf k}})+\textstyle{1 \over 2}} \right)} ,
\end{equation}
Similarly, in (\ref{eq14}),
\begin{equation}
\label{eq23}
\textstyle{1 \over 2}\left( {\nabla \phi } \right)^2\to \frac{1}{2V_s 
}\sum\limits_{\rm {\bf k}} {\frac{\vert {\rm {\bf k}}\vert ^2}{\omega _{\rm 
{\bf k}} }\left( {a^\dag ({\rm {\bf k}})a({\rm {\bf k}})+\textstyle{1 \over 
2}} \right)} 
\end{equation}
\begin{equation}
\label{eq24}
\textstyle{1 \over 2}m^2\phi ^2\to \frac{1}{2V_s }\sum\limits_{\rm {\bf k}} 
{\frac{m^2}{\omega _{\rm {\bf k}} }\left( {a^\dag ({\rm {\bf k}})a({\rm {\bf 
k}})+\textstyle{1 \over 2}} \right)} .
\end{equation}
Since $\left( {\omega _{\rm {\bf k}} } \right)^2=m^2+\vert {\rm {\bf 
k}}\vert ^2$, the above three relations summed reduce to the well-known 
operator form for (\ref{eq14})
\begin{equation}
\label{eq25}
\rho _\phi =\frac{1}{V_s }\sum\limits_{\rm {\bf k}} {\omega _{\rm {\bf k}} 
\left( {a^\dag ({\rm {\bf k}})a({\rm {\bf k}})+\textstyle{1 \over 2}} 
\right)} ,
\end{equation}
which has the non-zero vacuum expectation value (VEV)
\begin{equation}
\label{eq26}
\left\langle 0 \right.\vert \rho _\phi \vert \left. 0 \right\rangle =<\rho 
_\phi >=\frac{1}{V_s }\sum\limits_{\rm {\bf k}} {\frac{\omega _{\rm {\bf k}} 
}{2}} \quad .
\end{equation}
\subsection{Pressure of the Vacuum Fluctuations}
\label{subsec:pressure}
In similar fashion, the pressure operator of (\ref{eq15}) reduces to
\begin{equation}
\label{eq27}
p_{1\phi } =\frac{1}{V_s }\sum\limits_{\rm {\bf k}} {\frac{\vert k_1 \vert 
^2}{\omega _{\rm {\bf k}} }\left( {a^\dag ({\rm {\bf k}})a({\rm {\bf 
k}})+\textstyle{1 \over 2}} \right)} 
\end{equation}
and therefore from (\ref{eq25}) and (\ref{eq27})
\begin{equation}
\label{eq28}
\langle \rho _\phi \rangle \ne -\langle p_{1\phi } \rangle .
\end{equation}
In particular, on average, with most \textbf{k}, \textit{$\omega $}$_{k}>>m$, one finds
\begin{equation}
\label{eq29}
k_1^2 =k_2^2 =k_3^2 =\frac{\left| {\rm {\bf k}} \right|^2}{3}=\frac{\omega 
_{\rm {\bf k}}^2 -m^2}{3}\approx \frac{\omega _{\rm {\bf k}}^2 }{3}\quad .
\end{equation}
Thus, we have, from (\ref{eq25}) and (\ref{eq27}),
\begin{equation}
\label{eq30}
p_{\phi \,\mbox{flucts}} \approx \frac{\rho _{\phi \,\mbox{flucts}} 
}{3}\,\,\,\,\,\,\,\,\,w_{\phi \,\mbox{flucts}} \approx 
\frac{1}{3}\,\,\,\,\,\,\,\,\,\,\,\,\,\left( {\omega _{\rm {\bf k}} \gg m} 
\right)\quad .
\end{equation}
For lower energies, where $\omega _{\rm {\bf k}} \approx m$, (\ref{eq29}) becomes 
(using classical concepts with particle velocity \textbf{v} for 
illustration)
\begin{equation}
\label{eq31}
k_1^2 =\frac{\omega _{\rm {\bf k}}^2 -m^2}{3}=\textstyle{1 \over 3}\left( 
{\frac{m^2}{1-v^2}-m^2} \right)=\textstyle{1 \over 3}\left( {2m} 
\right)\left( {K.E.} \right)\ll \omega _{\rm {\bf k}} ^2,
\end{equation}
where the fourth part above is the standard ``low'' energy (still 
relativistic) case. For energies of this level, (\ref{eq27}) becomes much less than 
(\ref{eq25}), so effectively
\begin{equation}
\label{eq32}
p_{\phi \,\mbox{flucts}} \approx 0\,\,\,\,\,\,\,\,\,w_{\phi \,\mbox{flucts}} 
\approx 0\,\,\,\,\,\,\,\,\,\,\,\,\left( {\omega _{\rm {\bf k}} \approx m} 
\right).
\end{equation}
Thus the equation of state (\ref{eq4}) for scalar vacuum fluctuations has a range 
for $w$ of
\begin{equation}
\label{eq33}
\frac{1}{3}\ge w_{\phi \,\mbox{flucts}} \ge 0\quad .
\end{equation}
This is not the $w$ = --1 value of (\ref{eq7}), and thus these vacuum fluctuations 
cannot give rise to an effective cosmological constant, at any level of 
energy.

Similar results should be found for a complex scalar field, fermions, and 
spin 1 bosons.

\section{Conclusion}
\label{sec:conclusion}
When one calculates the vacuum fluctuations pressure from fundamental 
principles of QFT, in the same well accepted manner as the enormous vacuum 
fluctuation energy is calculated, one finds the pressure is not equal to the 
negative of the vacuum fluctuation energy density.

The common approach to vacuum fluctuation pressure determination comprises 
the following.

\begin{enumerate}
\item Assume vacuum energy density$\rho _{\mbox{vac}} $must be Lorentz invariant, constant in time, and give rise (in principle) to a cosmological constant.
\item Assume the $\raise.5ex\hbox{$\scriptstyle 1$}\kern-.1em/ \kern-.15em\lower.25ex\hbox{$\scriptstyle 2$} $ quanta energy summation in the VEV of the quantized Hamiltonian (equivalently, the time-time component of the quantized stress energy tensor), for example, $\rho _{\phi \,\mbox{flucts}} $, is vacuum energy obeying 1.
\item Conclude that one must then have vacuum fluctuation pressure $p_{\phi \,\mbox{flucts}} =-\rho _{\phi \,\mbox{flucts}} $ (i.e., $w_{\phi \,\,\mbox{flucts}} =-1)$. 
\end{enumerate}
On the other hand, the fundamental theoretic approach, which is generally 
accepted as valid for determining $w$ for the symmetry breaking vacuum energy 
density remnant, applied to vacuum fluctuations, comprises the following.

\begin{enumerate}
\item Use the quantized stress-energy tensor for a given field to calculate $\raise.5ex\hbox{$\scriptstyle 1$}\kern-.1em/ \kern-.15em\lower.25ex\hbox{$\scriptstyle 2$} $ quanta pressure, for example, $p_{\phi \,\,\mbox{flucts}} $.
\item Compare the value in 1 to the known $\rho _{\phi \,\,\mbox{flucts}} $.
\item Conclude that $0<p_{\phi \,\,\mbox{flucts}} <\frac{\rho _{\phi \,\,\mbox{flucts}} }{3}\,\,\,\,\,\,\,\left( {\mbox{i.e.,}\,\,w_{\phi \,\,\mbox{flucts}} \ne -1} \right)$.
\end{enumerate}
The common approach seems to ``put the cart before the horse'' by 
encompassing \textit{a priori} assumptions, whereas the fundamental theoretic approach starts 
from elemental principles, and does not. Thus, it is submitted that, 
contrary to common belief, vacuum fluctuations cannot qualify as a candidate 
for an effective cosmological constant (which requires $w$ = -- 1), regardless 
of order.

This can mean one of two things. For one, an undetermined symmetry 
\cite{See:2} may exist that cancels out the $\raise.5ex\hbox{$\scriptstyle 
1$}\kern-.1em/ \kern-.15em\lower.25ex\hbox{$\scriptstyle 2$} $ quanta 
contributions leaving a net vacuum fluctuation energy of zero (which is 
Lorentz invariant, constant, and has no cosmological implications). 
Alternatively, and less satisfying, the $\raise.5ex\hbox{$\scriptstyle 
1$}\kern-.1em/ \kern-.15em\lower.25ex\hbox{$\scriptstyle 2$} $ quanta energy 
density may simply be neither Lorentz invariant nor constant in time.

\section*{Acknowledgement}
I would like to thank P. J. E. Peebles for reading the manuscript and agreeing that vacuum 
fluctuation $w\ne $ --1.

\end{document}